\documentclass[a4paper,12pt]{article}
\usepackage[warn]{mathtext}
\usepackage[T2A]{fontenc}
\usepackage[cp1251]{inputenc}
\usepackage[english]{babel}
\usepackage{amssymb,amsfonts,amsmath,mathtext,cite,enumerate,float}
\usepackage{graphicx}
\usepackage{indentfirst}

\makeatletter
\renewcommand{\@biblabel}[1]{#1.}
\makeatother

\usepackage{geometry}
\begin{document}

\begin{center}
\textbf{Correlation effects during liquid infiltration into hydrophobic nanoporous mediums}
\end{center}

\begin{center}
\textbf{V.D. Borman$^1$, A.A. Belogorlov$^1$, V.A. Byrkin$^1$, G.V. Lisichkin$^2$, V.N. Tronin$^1$, V.I. Troyan$^1$}
\end{center}

$^{1}$Department of Molecular Physics, Moscow Engineering Physics Institute, Moscow 115409, Russia\\ 
$^{2}$Department of Chemistry, M. V. Lomonosov Moscow State University, Moscow 119992, Russia

\begin{abstract}
Correlation effects arising during liquid infiltration into hydrophobic porous medium are considered. On the basis of these effects a mechanism of energy absorption at filling porous medium by nonwetting liquid is suggested. In accordance with this mechanism, the absorption of mechanical energy is a result expenditure of energy for the formation of menisci in the pores on the shell of the infinite cluster and expenditure of energy for the formation of liquid-porous medium interface in the pores belonging to the infinite cluster of filled pores. It was found that in dependences on the porosity and, consequently, in dependences on the number of filled pores neighbors, the thermal effect of filling can be either positive or negative and the cycle of infiltration-defiltration can be closed with full outflow of liquid.  It can occur under certain relation between percolation properties of porous medium and the energy characteristics of the liquid-porous medium interface and the liquid-gas interface. It is shown that a consecutive account of these correlation effects and percolation properties of the pores space during infiltration allow to describe all experimental data under discussion.
\end{abstract}

\section{INTRODUCTION}

Energetics of ``nanoporous medium--nonwetting liquid'' systems is one of the new directions in basic and applied research (see e.g. \cite{1,2,3,4,5,6,7,8,9,10,11,12}). In the simple model of a porous medium in the form of cylindrical channels, threshold pressure is described by the Laplace-Washburn equation $p=2\sigma /R \left| {cos\alpha } \right|$, where $\sigma $ is the surface energy of the liquid, $R$ is the pore radius, and $\alpha $ is the contact angle (for a nonwetting liquid, $\alpha >90^\circ$). For filling nanometer-sized pores by a nonwetting liquid with a surface energy of $0.05\div 0.5$ J/m$^{2}$, this threshold pressure is $10^2\div 10^3$ atm. When the liquid passes from the bulk state to a dispersed state in pores of the nanoporous medium with a specific volume of $\sim 1$ cm$^{3}$/g, the absorbed and accumulated (returned when the liquid flows out) energy can reach $10\div 100$ kJ/kg. This value is one order of magnitude higher than the energy observed for widely used materials, such as polymer composites and alloys with the shape memory effect \cite{10,11,12}. 

Among the systems under investigation are silochromes, zeolites with liquid metals, hydrophobized silica gels, and zeolites with water and aqueous solutions of organic compounds and salts. In recent years, hydrophobized nanoporous media have become available owing to the development of the method used for modifying the surface of nanoporous media, generally, with alkyl chlorosilanes \cite{3,6,7,12,13,14,15,16,17,18,19,20}. To date, nanoporous media with different pore shapes, porosities, specific surface areas, specific volumes, average pore radii, and pore size distributions have been studied \cite{1, 2, 3, 4, 5, 6, 7, 17 , 18, 19 , 20, 21, 22, 23, 24, 25, 26, 27, 28, 29, 30, 31, 32, 12, 34, 35, 36, 10, 11, 39, 40, 41, 42, 43, 44, 45, 46, 47, 48, 49, 50, 51, 9, 53, 55}. The investigations performed thus far have been concerned primarily with equilibrium properties. Experiments have been carried out at a low compression rate of the system when the rate of increase in the pressure ($\dot {p}$) in the liquid-porous medium system is ($10^{-3}-1$) atm/s. In the infiltration-defiltration cycle, there is a hysteresis, so that the threshold pressure of infiltration is higher than the pressure of defiltration. Moreover, the majority of the  systemsare characterized by the phenomenon of nonoutflow of a nonwetting liquid when a part of this liquid remains in the porous medium as the excess pressure decreases to zero. The absorbed energy is determined by the product of the volume of filled pores and the difference between the infiltration and defiltration pressures. In frameworks the model of cylindrical channels, these pressures are described by the Laplace-Washburn equation with different contact angles.

The revealed difference between the infiltration and defiltration pressures and the absorption of the mechanical energy observed in the infiltration-defiltration cycle due to the pressure hysteresis, as a rule, have been explained by the hysteresis of the contact angle; however, the mechanism responsible for the appearance of the latter hysteresis has remained unclear \cite{4,6,7,19,20,21,22}. 

It has been established that the infiltration and defiltration pressures depend on the temperature and that, for the porous medium with a disordered structure of pores, the defiltration pressure increases (by several factors) with an increase in the temperature from 280 to 350$^{\circ}$K, whereas the infiltration pressure decreases only slightly (by less than 10{\%}) or remains constant \cite{21,22,24}. This means that, during infiltration and defiltration, the phenomenological contact angles differently depend on the temperature. For zeolites, the revealed temperature dependences exhibit a more complex behavior; moreover, the volume ($V$) memory effect can be observed with an increase in the temperature and its subsequent decrease. It is worth noting that this effect is one order of magnitude (in $\Delta V/V$) stronger than that observed for known alloys and composites \cite{10,11}. 

At present, there exist several hypothesis regarding the nature of the contact angle hysteresis. This hysteresis has been attributed to the rough surface of pores, the chemical inhomogeneity of the surface, and the dependence on the direction of the liquid motion \cite{19,20}. 

However, these conceptions cannot explain the different dependences on the temperature of infiltration and defiltration pressures, and hence the temperature dependence of contact angle.

In the framework of the concept that the porous medium is a system of cylindrical channels, the absorbed energy is expended for forming a liquid-porous medium surface, which appears in the course of infiltration and disappears during defiltration at different pressures due to different contact angles \cite{6,7,19, 20, 21, 22,49,50,2}. In case of the closed hysteresis loop in the infiltration - full-defiltration process initial and finite states of the system are similar, the internal energy change $\Delta E_{cicle} =\oint {dE=0} $ and the work done to perform the filling of the porous medium should be equal to the thermal effect $\oint {dA=\oint {dQ} } $. The measurements carried out in \cite{19,58} showed that rise of temperature in the hydrophobic silica gel-water systems under investigation during infiltration-defiltration cycle was $<$10$^{-3}$K. On the other hand, it was found in \cite{46} that when one-third of the porous medium volume of a similar system was filled and the liquid-porous medium interface area change was not equal to zero, the temperature did not increase within the limits of error ($\le 0.1$K), while the estimation provides the temperature increase by $\Delta T = 0.8$K.

In \cite{2} the dependence of the thermal effect $\Delta Q$, which is accompanied by the thermal effect due to elastic compression of water and the porous medium, on the filling degree in the hydrophobic silica gel-water system was measured. The author found that during infiltration this porous medium with $\varphi \approx 0.22$, the heat generation took place ($\Delta Q<0$), whereas $\frac{d\sigma }{dT}<0$ for water \cite{58} and, consequently, in case of independence of the contact angle from temperature during the formation of liquid-solid interface with area $\Delta S$ one would expect the $\Delta Q=-\frac{d\sigma }{dT}\Delta S$ value to be positive, i.e. heat absorption shall be observed. In \cite{2}, it was experimentally found that, as the degree of infiltration increases, the quantity $\left| \Delta Q \right|$ reaches a maximum at $\theta \approx 0.6$ and decreases to zero after the complete infiltration of the porous medium: $\Delta Q=0$. This result led the author of \cite{57} actually to the conclusion that the first law of thermodynamics is violated in the process under investigation.

Thermal effect associated with the infiltration of the modified porous medium was also observed \cite{50,51}. It was found that for the investigated porous media with $ \varphi =0.4 $ and $\varphi =0.6$ the value $ \Delta Q <0 $ reduces with increasing degree of filling, reaching a minimum at complete filling.

Note that the energy absorption can not be to all appearance explained by viscous dissipation, since, as it ascertained in [59], change in viscosity of nonwetting liquid (aqueous solution of CaCl$_{2}$) by 7 times does not alter the dependence of the filled volume on time, the threshold infiltration pressure and rate of filling hydrophobic silicagel.

Statistical theory of fluctuations allows us to generalize the Laplace-Washburn equation in case of filling the pores in the system of interconnected pores \cite{4,5}. The proposed approach makes it possible to analyse the probability of the fluctuation infiltration pores with the size $R$ near the threshold pressure. If the probability of $w\approx \exp (-\delta A/T)$ increases continuously with growing pressure, then the pore can be filled with liquid ($\delta A$ is work expended for infiltration of one pore, $\delta A(p,\sigma ,\delta \sigma ,R)$). The value $\delta A_{in} $ takes into account a contribution of formation a liquid-porous medium interface and a surface of menisci at the mouths of the filled pores. Threshold infiltration pressure is determined from the condition $\delta A_{in} (p_{out} ,\sigma ,\delta \sigma ,R)=0$, and the defiltration pressure is determined from the condition $\delta A_{out} (p_{out} ,\sigma ,\delta \sigma ,R)=0$. This makes it possible to calculate contact angles during infiltration and defiltration, to describe the dependences of the volume of system liquid-porous medium on the pressure and calculate the volume of confined liquid after defiltration.

In \cite{60} the energy absorption during the infiltration-defiltration process within the framework of fluctuation model is associated with the energy of formation of menisci. However, in this model is assumed, that the work of filling the pores (liquid defiltration) and, consequently, the probability of infiltration (defiltration) does not depend on the degree of filling of the porous medium. Therefore, if during infiltration and defiltration  the contact angle is independent of the filling degree, the area of menisci which are formed should be equal to the area of menisci disappeared, and therefore the formation energy at full infiltration (defiltration) should be equal to zero. It also does not allow to explain the results of \cite{49,50,2}.

Thus, the experimental data currently known seems to be self-contradictory and well-known traditional mechanisms of energy absorption do not make it possible to explain the infiltration-defiltration hysteresis, temperature dependences of the infiltration and defiltration pressures and special features of heat generation during the filling of nanoporous media with nonwetting liquid which are observed.

As shown in this paper (Section 2), to describe the discussed phenomena we should take into account the structure of pores in disordered porous medium. 

During infiltration into porous medium with an increase in the number of filled pores, surrounded by empty filled pores a formation meniscus in pore mouths is changed by theirs disappearance on the average ensemble of pores with increasing number of full filled pores surrounded by empty loadable pores. In this case expenditure of work on formation meniscus are replaced in the gain energy during theirs disappearance. Thus, the correlation effect of the relative position of filled and empty pores must be taken into account at describing the fluctuation infiltration. In addition, the porosity of the medium determines the possible neighbors number of pores and possible macroscopic set of filled pores in the percolation cluster, through which liquid flow from the surface of the porous medium to fill the pores is possible. This means necessity of taking into account another correlation effect -- the spatial arrangement of pores in the medium, arising as a result of probabilistic system realization of interconnected pores in the infinite percolation cluster.	
	
Consequently, it is possible to sugges the mechanism of energy absorbtion during filling of the nanoporous medium with nonwetting liquid is proposed (Sections 3, 4). In accordance with this mechanism, the mechanical energy absorption is a result of expenditure of energy for the formation of menisci in the pores on the shell of the infinite cluster and expenditure of energy for the formation of liquid-porous medium interface in the pores belonging to the infinite cluster of filled pores. It has been demonstrated (Section 4), that the infiltration-defiltration cycle can be closed with the complete defiltration of the liquid and the reproducibility of the cycle when the specific relationship between the percolation properties of the porous medium and the energy characteristics of the liquid-porous medium and liquid-gas interfaces is satisfied. It turned out that, depending on porosity and, consequently, the number of nearest neighbors of the filled pores, the thermal effect can be either positive or negative. This makes it possible to explain the known experimental data \cite{46,49,50} mentioned above and to describe the temperature dependences of the infiltration and defiltration pressures \cite{21,22,24}. The description of experimental data on the basis of the proposed approach is given in sections 5, 6.

\section{THE MODEL OF A POROUS MEDIUM. INFILTRATION FLUCTUATIONS}

Let us consider a disordered porous medium infiltrated with a nonwetting liquid. It is assumed that the half-width of the pore size distribution $\delta R$ satisfies the inequality $\delta R/\bar {R}<3$, so that the fulfillment of this inequality ensures the independence of the percolation threshold from the radii of pores \cite{61}. 

\begin{figure}[h]
\center{\includegraphics[width=0.5\linewidth]{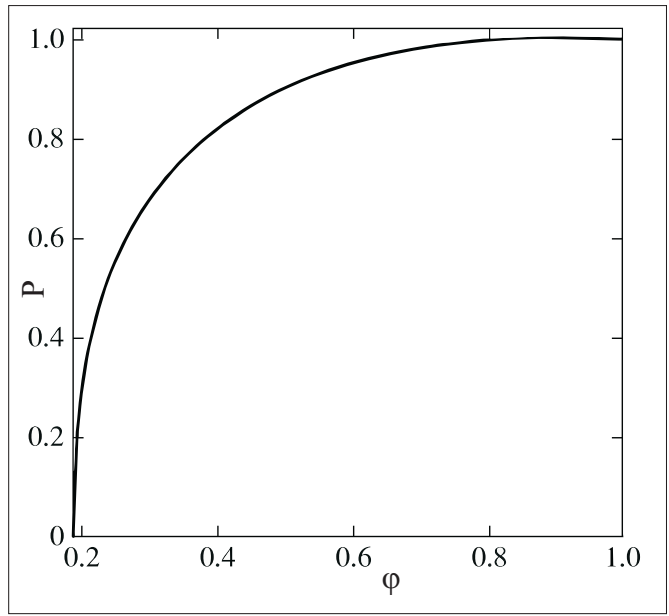}}
\caption{The dependence of the probability normalized to unity of a pore belonging to the infinitely large cluster on porosity $\varphi$}
\label{fig1}
\end{figure}

Filling of the porous medium is a process of liquid infiltration into the disordered porous medium which contains pores of different size. It is assumed that the size of the porous medium $a$ is much more than the maximum size of the pores $R_{\max }$ ($a\ge 10^3R_{\max }$ \cite{59}) so that the porous medium can be regarded as infinite. Obviously, the infiltration of all the pores can take place only when the pores are connected with the surface and form a connected system. Consequently, the filling of the porous medium can take place only when the pore system in it is far beyond the percolation threshold ($\varphi >\varphi _c $), $\varphi $ is its porosity equal to the ratio of pores to the porous medium volume, $\varphi _c $ is the percolation threshold, which is the characteristic of the porous medium. For 3D systems, the percolation threshold $\varphi _c =0.18$ \cite{62}. At the same time, the connectivity of pores with one another is the result of the formation of infinitely large clusters of pores at $\varphi =\varphi _c $. Figure~\ref{fig1} shows the dependence of the probability normalized to unity of a pore belonging to the infinitely large cluster on porosity $\varphi$ \cite{62,63}.

It can be seen from Fig.~\ref{fig1} that, in the vicinity of the percolation threshold $\varphi _c$, only a small number of pores ($\sim 1${\%}) belong to the infinite cluster; therefore, in this case, only a small fraction of these pores, as well as pores that belong to the finite clusters connected with the boundary of the porous medium, can be infiltrated. At increasing porosity and for $\varphi \gg \varphi _c  \quad P(\varphi )\to 1$ and, consequently, the pore space becomes homogeneous due to the growth of the infinitely large cluster of pores. Under these conditions, the infiltration of the porous medium can be described as the infiltration of an infinite cluster of pores. It is this infiltration that will be considered in the present paper. 

It is assumed that thermal fluctuations at the pressure $p$ in the vicinity of the infiltration threshold of the porous medium bring about the formation of macroscopically small regions in the form of clusters consisting of $N$ pores filled with a liquid. Each cluster arises at the boundary of the porous medium and, in the view of the boundedness of the pore volume, grows through the attachment of the other filled pores to it. We believe that, at the beginning of the growth, each cluster can be considered a system of branched chains consisting of filled pores. In the course of infiltration of the porous medium with the liquid, the external pressure does the work. This process is accompanied by the formation of energetically unfavorable surfaces of both the menisci of the liquid in pores and the liquid-porous medium interfaces. Moreover, the state of the gas in the pores and the elastic state of the porous medium change as well. If the adiabatic work of formation of an infiltration fluctuation is $\delta A(N)$ and the energy of dissipation due to the friction can be disregarded (see \cite{59}), the probability of the formation of a fluctuation can be written as $w\sim w_0 exp(\delta S)$ \cite{64}, where $\delta S=-\delta A/T$ is the fluctuation of the entropy. Therefore, an increase in the quantity $\delta A$ ($\delta A>0$) with an increase in the number of pores $N$ in the cluster leads to a decrease in the fluctuation probability. This corresponds to the thermodynamic stability of the initial state of the system. For $\delta A \sim T$, the infiltration fluctuation can increase. In this case, the system becomes unstable and the liquid begins to infiltrate the porous medium. 

The infiltration of a nonwetting liquid under the pressure $p_{in}$ a porous medium requires a certain amount of work to be done to fill the pores of the porous medium. For this purpose, it is necessary to overcome a certain critical pressure, which is the Laplace pressure $p_c (R)\sim \frac{\delta \sigma }{R}$ for an isolated pore with characteristic size $R$. $\delta \sigma =\sigma ^{sl}-\sigma ^{sg}$, where $\sigma ^{sl}$ is the surface energy of the solid-liquid interface, $\sigma ^{\lg }=\sigma $ is the surface energy of the solid-gas interface, $\delta \sigma =\sigma ^{sl}\left| {\cos \alpha } \right|$, $\alpha $ is the contact angle.

At a pressure lower than the critical value $p_{in}$, the adiabatic work satisfies the inequality $\delta A(N)>0$ at any value of $N$ and the fluctuation probability decreases with an increase in $N$. Therefore, the fluctuation probability is equal to zero for any macroscopically large number of pores. Fluctuations of finite length arise, but no infiltration of the porous medium occurs. At a pressure in the vicinity of the critical value $p_{in}$, the work is $\delta A \sim T$ and thermal fluctuations in the system can lead to the formation of clusters from $N$ pores. At a pressure $p>p_{in}$, the infiltration of individual pores becomes energetically favorable because the quantity $\delta A(N)$ is negative. Since the work is $\delta A\sim N$, the fluctuation probability at a pressure $p>p_{in}$ is $w\sim 1$. The pressure difference $p-p_{in}$ causes the liquid to move in the porous medium. 

Now, we consider a porous medium immersed in a nonwetting liquid under an external pressure $p$, which does the work in the course of infiltration of the porous medium. Let $\delta A\left( p \right)$ be the work expended for providing the fluctuation infiltration of one pore. According to \cite{5}, the expression for the work $\delta A\left( p \right)$ with due regard for the formation of menisci can be written in the form:

\begin{equation}
\label{eq1}
\delta A\left( p \right)=-pV+\sigma S_m +\delta \sigma (S-S_m ).
\end{equation}
Here, $V$ is the volume of the pore, $S$ is the surface area of the pore, $S_m$ is the surface area of the menisci, and $\sigma $ is the surface energy of the liquid. 

For a spherical pore with the radius $R$, the work $\delta A\left( p \right)$ can be represented in the form \cite{5}: 

\begin{equation}
\label{eq2}
\begin{gathered}
\delta A =A\left( {p,R} \right)\frac{4 \pi R^3}{3}, \\
A \left( {p,R} \right)=-p+\frac{3 \delta \sigma }{R} \left[ {1+\eta \left( {\frac{\sigma }{\delta \sigma }-1} \right)} \right],
\end{gathered}
\end{equation}
where $\eta $ is the ratio of the meniscus surface area to the pore surface area. 

A similar expression can be written for a cylindrical pore. For example, we write the following expression for the work expended for providing the fluctuation emptying of a cylindrical pore by nucleation with the nucleus length $L$ and the radius $R$: 

\begin{equation}
\label{eq3}
\begin{gathered}
\delta A =A\left( {p,R,L} \right) 4/3 \pi R^3,\\
A \left( {p,R,L} \right)=p(\frac{3x}{2}-1)-\frac{3 \delta \sigma }{R} \left[ {x-\frac{\sigma }{\delta \sigma }} \right],
\end{gathered}
\end{equation}
where $x=L/2R$.

It follows from relationship (\ref{eq2}) that the infiltration of the pore becomes energetically favorable at a pressure higher than the critical value $p_{in}$. The critical pressure is determined by the equality of the work on the fluctuation infiltration of the pore to zero. An analysis of the critical pressure for a spherical pore was performed in \cite{5}. Expression (\ref{eq3}) implies that, devastation of the pores becomes energetically favorable at a pressure less than the critical pressure $p_{out} $, which depends on the nucleus length $L$ and the nucleus radius $R$ and, at $L\to \infty $, transforms into the Laplace pressure $p_{out} \sim \frac{2\delta \sigma }{R}$. This means that, in the work expended for infiltrating the pore, the change in the surface energy of the pore dominates over the energy expended for forming menisci.

For characteristic values of the parameters of the porous medium and the liquid $\sigma \sim \delta \sigma \sim 0.5$ kJ/m$^{2}$, $R\sim 10$ nm, and, $\chi \sim 10^{-4}$ atm$^{-1}$, when the pressure deviates from the value $p_{I}$ by $\delta p=-10^{-2} p_{I}$, the work changes from $\delta A\sim T$ to $\sim 1$ eV. In this case, the change in the probability of infiltration fluctuation at the temperature $T = 400$K is equal to ten orders of magnitude. Therefore, for systems with the aforementioned characteristic parameters, the probability $w$ changes abruptly from 0 to 1 in a narrow pressure range ($\delta p/p \sim 10^{-2}$) in the vicinity of the pressure $p_{in}$. It should be noted that the inclusion of the gas filling the pores in the analysis leads to the appearance of an additional contribution to the work $\delta A_g$. The value of this contribution $\delta A_g $ during filling one pore with the volume $V$ can be estimated under the assumption that it is equal in the order of magnitude to the specific evaporation energy of the liquid $\mu _0 $ multiplied by the density of the gas $\rho _g $ at the pressure $p$, i.e., $\delta A_g \sim \mu _0 \rho _g V$. For water, we can write $\mu _0 \sim 2.2\cdot 10^3$ J/kg and $\rho _0 \sim 1.0$ kg/m$^3$. Assume, that the filling of the porous medium occurs at a pressure of the liquid $p_{in} \sim 1 \cdot 10^7$ J/m$^3$. In this case, the additional contribution to the work due to the presence of the gas in the pores is given by $\delta A_g \sim \mu _0 \rho _g V\sim 10^{-2} \cdot \delta A$. Nominally, accounting of the influence of gas reduces to the appearance of additional summands in equations (\ref{eq2}) and (\ref{eq3}), each of them proportional to the gas pressure at a given temperature. The value of these summands $A_{gin}$ and $A_{gout}$ for typical infiltration pressure $p_{in} \sim 200$ atm and $p_{out} \sim 1-10$ atm, for pore radius $R \approx 10$ nm is $A_{gin} \sim A_{g} \sim A_{gout} \sim 10^{-2}$eV in order of magnitude. Therefore, the influence of the gas in the pores on the infiltration of the porous medium can be ignored  at considering the filling of the pore, but accounting of these influence may be important at considering the effects of the defiltration liquid the cylindrical pores. 

Since the pore can be either filled (probability $w\sim 1$, $\delta A\left( p \right)<0)$ or empty (probability $w=0$, $\delta A\left( p \right)>0$), the normalized probability can be written in the form \cite{5}

\begin{equation}
\label{eq4}
w_i \left( p \right)=\left[ {1+\exp \left( \delta A / T\right)} \right]^{-1}.
\end{equation}

Note, that relations (\ref{eq3}) and (\ref{eq4}) explain the obtained in \cite{6,7} experimental data for the dependence of infiltration pressure and defiltration pressure on the pore size.

Pores are not isolated in a porous medium, but they are connected with one another by throats (mouths), in which menisci are formed during infiltration of a certain pore. Thus we can assume that the medium subjected to infiltration is the heterogeneous medium which consists of full and empty pores playing the role of white and black spheres, respectively, in the percolation theory \cite{62}. Such medium can experience percolation transition occurring via the formation of clusters of accessible pores and followed by infiltration of nonwetting liquid into such formations. \cite{59}. In addition to that, filling of the macroscopic volume of a porous medium occurs by infiltration in the infinitely large cluster of accessible pores \cite{59}.

Below, we will consider the infiltration of pores located on the shell of an infinite cluster consisting of filled pores. In this case, the condition $\delta A=0$ determines the pressure necessary for the infiltration of a pore on the shell of the infinite cluster of filled pores.

\section{WORK AND THERMAL EFFECT IN THE INFILTRATION-DEFILTRATION CYCLE}

Let us calculate the work $\delta A$ and the thermal effect $\Delta Q$ for an arbitrary degree of infiltration $\theta $ of the porous medium. The thermal effect $\Delta Q$ in filling of a porous medium by nonwetting liquid comprises the thermal effect $\Delta Q_p $ due to the formation of the liquid-solid interface , the thermal effect $\Delta Q_w $ related to formation-disappearance of menisci and the thermal effect $\Delta Q_u $ related to the compressibility of the nonwetting liquid-nanoporous medium 
system.

\begin{equation}
\label{eq5}
\Delta Q=\Delta Q_p +\Delta Q_w +\Delta Q_u .
\end{equation}

The $\Delta Q_p $, $\Delta Q_w $ and $\Delta Q_u $ values can be calculated using thermodynamic relations \cite{64}, which determine the thermal effect in formation of the surface $\Delta Q_s $:

\begin{equation}
\label{eq6}
\Delta Q_s =-\frac{d\sigma }{dT}\Delta S.
\end{equation}

Here, $\Delta S$ is the change in the system surface.

To calculate the thermal effect $\Delta Q_s $ let us suppose that each pore in a porous medium has $z$ nearest neighbours and pores contact each other by throats, each of which has an area $S_z$. If an empty pore contacts a full one, the meniscus is formed in the throat. A pore in a porous medium can be filled only if liquid can reach it. In compliance with above-mentioned assumptions this condition can be satisfied by formation of an infinitely large cluster of filled pores. In this case, only those pores which belong to the shell of the infinite cluster will be filled. It is possible to show that the contribution of filled finite-size clusters (which liquid can reach via filled clusters contacting with the surface of the porous medium)to the filled volume is small. Distribution $f(N)$ of the number of pores in clusters of finite size near the percolation threshold is determined by the scaling dependence $f(N)\sim \frac{1}{N^\tau }$, $\tau =2.2$ \cite{62}. Therefore the bulk of the cluster contains one or more pores, which are mostly not associated with the surface of the porous medium. Liquid cannot reach such pores and, consequently, they are not filled at $\theta \sim \theta _c $. Taking it into consideration, we can represent the thermal effect $\delta Q$ related to filling of one pore as:

\begin{equation}
\label{eq7}
\delta Q=-T\frac{d\delta \sigma }{dT}(S-zS_z )-T\frac{d\sigma }{dT}zS_z W(\theta ),
\end{equation}
$S=4\pi R^2$ is the area of the surface of a pore with radius $R$, $z$ is a number of nearest neighbours, $W(\theta )$ is the difference (averaged over the ensemble of pores) between the numbers of menisci before and after the infiltration of the pore per the nearest neighbor of the infinite cluster.

Considering that filling of a porous medium is the result of formation of the infinitely large cluster of filled pores and taking into account the normalized probability $P(\theta )$, we obtain that the quantity of heat per a pore released in the process of filling the porous medium to the degree of filling $\theta $ can be written as:

\begin{equation}
\label{eq8}
\begin{gathered}
 \Delta Q_p (\theta )=-T\frac{d\delta \sigma }{dT}\int\limits_0^\theta {<(S-zS_z )>\theta P(\theta )d\theta }, \\ 
 <S-zS_z >=\int\limits_0^\infty {dR} (S-zS_z )f(R), \\ 
	\Delta Q_w (\theta )=-T\frac{d\sigma }{dT}\int\limits_0^\theta {<zS_z W(\theta )>d\theta }.
\end{gathered}
\end{equation}

Here, $f(R)$ is the function normalized to unity of the size distribution for pores. For a disordered porous medium average values in (\ref{eq8}) can be calculated in the framework of a specific model of a porous medium. We will use the model of randomly arranged spheres in which pores represent randomly arranged spherical holes \cite{65}. This model does not take into account correlations in location of pores with different radii in accordance with assumption made about the narrowness of size distribution for pores $(\delta R) \ll \bar {R}$. In accordance with the model, the average number of nearest neighbours $\bar {z}$, associated with porosity of medium $\varphi $, and the area of a throat can be written in forms \cite{65}:

\begin{equation}
\label{eq9}
\begin{gathered}
z=\bar {z}=-8\ln (1-\varphi ),\\
S_z =\frac{9\pi ^2}{256}\bar {R}^2.
\end{gathered}
\end{equation}

Using expression (\ref{eq8}) and taking into account that $(\delta R)\ll \bar {R}$, we obtain from (\ref{eq9}):

\begin{equation}
\label{eq10}
\begin{gathered}
 \Delta Q_p (\theta )=-T\frac{d\delta \sigma }{dT}(1-\eta )4\pi \bar {R}^2\int\limits_0^\theta {\theta P(\theta )d\theta }, \\ 
 \Delta Q_w =-T\eta \frac{d\sigma }{dT}4\pi \bar {R}^2\int\limits_0^\theta {W(\theta )d\theta }, \\ 
	\eta =\bar {z}\frac{9\pi }{1024}^2.
\end{gathered}
\end{equation}

To calculate $W(\theta )$ we consider an empty pore located on the perimeter of the infinite cluster of filled pores. Let us suppose that this pore contacts the infinitely large cluster of filled pores via n throats. Thus, menisci are formed in all $n$ mouths and menisci are absent in the remaining $z-n$ throats. After filling this pore, menisci which were there at the beginning of infiltration disappear and the number of menisci will be equal to $z-n$. In this case we can write $W(\theta )$ as:

\begin{equation}
\label{eq11}
W(\theta )=\sum\limits_{n=1}^z {(P(\theta )})^n(1-\theta)^{z-n+1}\frac{z-2n}{z}\frac{z!}{n!(z-n)!}.
\end{equation}

The first factor under the summation sign determines the probability that an empty pore contacts the infinite cluster of filled pores n times, the second factor is the probability of finding the empty pore close to the infinite cluster, provided that this pore is surrounded by $z-n$ empty pores and therefore has $z-n$ throats. The third factor determines the difference between a relative number of menisci after ($z-n$) and before ($n$) the filling of the pore. The binomial coefficient takes into account variants of allocation of $n$ menisci on number of pore nearest neighbours. Note that the obtained expression coincides with the full perimeter of the infinitely large cluster calculated in \cite{66}, if the third factor is substituted for unity.

The sum in Eq. (\ref{eq11}) can be calculated analytically:

\begin{equation}
\label{eq12}
W(\theta )=(\theta ^2-2\theta -P(\theta )+1+\theta P(\theta ))(P(\theta )-\theta +1)^{z-1}-(1-\theta )^{z+1}.
\end{equation}

Relations (\ref{eq9}), (\ref{eq10}), (\ref{eq12}) determine the thermal effect during filling a porous medium with porosity $\varphi $ to the degree of filling $\theta $.

The work $\Delta A$ expended for infiltrating the porous medium to the volume determined by the fraction $\theta $ and the corresponding work expended for infiltrating one pore $\delta A_{in} (\theta )$ can be calculated from the thermodynamic relationship $A=\int \sigma dS$ \cite{64}. By using expression (\ref{eq2}) for spherical pores, we obtain:

\begin{equation}
\label{eq13}
\begin{gathered}
 \delta A_{in} (\theta ,p)=-p\frac{4\pi \bar {R}^3}{3}+\delta \sigma (1-\eta )4\pi \bar {R}^2+\sigma \eta 4\pi \bar {R}^2W(\theta ) ,\\ 
 \Delta A=\Delta A_p +\Delta A_w ,\\ 
 \Delta A_p (\theta )=\delta \sigma (1-\eta )4\pi \bar {R}^2\int\limits_0^\theta {\theta P(\theta )d\theta } ,\\
 \Delta A_w =\sigma \eta 4\pi \bar {R}^2\int\limits_0^\theta {W(\theta )d\theta }
\end{gathered}
\end{equation}

The sum of heat (\ref{eq10}) and work (\ref{eq13}) determines the change in the energy of isothermal infiltration of the porous medium:

\begin{equation}
\label{eq14}
\begin{gathered}
 \Delta E=\Delta E_p +\Delta E_w,\\ 
 \Delta E_p =(\delta \sigma -T\frac{d\delta \sigma }{dT})(1-\eta )4\pi \bar {R}^2\int\limits_0^\theta {\theta P(\theta )d\theta },\\
 \Delta E_w =(\sigma -T\frac{d\sigma }{dT})\eta 4\pi \bar {R}^2\int\limits_0^\theta {W(\theta )d\theta }. \\ 
\end{gathered}
\end{equation}

It follows from (\ref{eq14}) that the change in energy of the system during filling the porous medium is determinded by the specific surface energies $\sigma $ and $\delta \sigma $, geometric properties of the porous medium and the evolution of the infinite cluster of filled pores, which depends on the properties the disordered porous medium (Fig.~\ref{fig1}). 

For the calculation of the work $\delta A(\theta )$ and the thermal effect $\Delta Q_v$ arising upon the defiltration of the liquid from the porous medium, it should be noted that, in the infiltrated porous medium, the defiltration of the liquid leads to the formation of empty pores surrounded by at least one filled pore connected through other filled pores with the surface of the porous medium. As in case of filling, the formation of an empty pore goes with the change in surface energy of the liquid-solid and liquid-gas interfaces as well. The change in surface energy is associated with formation-disappearance of menisci \cite{5}.

Taking into account this fact, the work expended for emptying the pore in the porous medium $\delta A(\theta )$ with the degree of infiltration $\theta $, the work for defiltration liquid $\Delta A^v$ from $1$ to $\theta $ the degree of infiltration, and the thermal effect $\delta Q_v $ associated with the defiltration of the liquid from one pore in the porous medium can be written in the form:

\begin{equation}
\label{eq15}
\begin{gathered}
	\delta A_{out} (\theta )=p\frac{4\pi \bar {R}^3}{3}-\delta \sigma (1-\eta )4\pi \bar {R}^2+\sigma \eta 4\pi \bar {R}^2W_1 (\theta ),\\
	\Delta A^v_p (\theta )=\delta \sigma (1-\eta )4\pi \bar {R}^2\int\limits_0^\theta {\theta }'d{\theta }', \\
	\Delta A^v_w =-\sigma \eta 4\pi \bar {R}^2\int\limits_0^\theta {W_1 ({\theta }')d{\theta}'} ,\\
	\delta Q_v =-T\frac{d\delta \sigma }{dT}(S-zS_z )+T\frac{d\sigma }{dT}zS_z W_1 (\theta ).
\end{gathered}
\end{equation}
Relations (\ref{eq7}) and (\ref{eq15}) differ in sign of the last term and the functions $W(\theta )$ and $W_1 (\theta )$ which determine the difference per one nearest neighbour between the number of menisci before and after infiltration (defiltration) in pores.

In contrast to the case of infiltration, the defiltration of the liquid occurs initially through the formation of individual empty pores and clusters of empty pores with a decrease in the pressure and, after the infinite cluster of empty pores is formed, through the formation of pores on the shell of this cluster. Upon the defiltration when the low degree of infiltration is reached, the liquid can be retained in the porous medium if it is contained in individual pores or clusters of filled pores surrounded by empty pores with smaller sizes from which the liquid defiltrated at higher pressures. However, as the number of neighbors of empty pores increases, according to relationships (\ref{eq15}) (see below Figs.~\ref{fig1}-\ref{fig3}), the defiltration of the liquid becomes energetically more favorable; i.e., it should proceed at higher pressures. Therefore, the quantity $W_1 (\theta )$ should be defined as the difference (averaged over the ensemble of pores) between the numbers of menisci before and after the emptying of the pore on the shell of the system of empty pores. Taking it into consideration, calculation of $W_1 (\theta )$ gives:

\begin{equation}
\label{eq16}
W_1 (\theta )=(2\theta ^2-3\theta +1)-(1-\theta )^{z+1}.
\end{equation}
The thermal effect, the work, and the change in the energy during the defiltration, when the degree of infiltration varies from $1$ to $\theta $, can be written in the form similar to relationships (\ref{eq10}), (\ref{eq13}), and (\ref{eq14}):

\begin{equation}
\label{eq17}
\begin{gathered}
	\Delta Q^v_p (\theta )=-T\frac{d\delta \sigma }{dT}(1-\eta )4\pi \bar {R}^2\int\limits_0^\theta {\theta d\theta } ,\\
	\Delta Q^v_w =T\eta \frac{d\sigma }{dT}4\pi \bar {R}^2\int\limits_0^\theta {W{ }_1(\theta )d\theta }, \\
	\Delta A^v_p (\theta )=\delta \sigma (1-\eta )4\pi \bar {R}^2\int\limits_0^\theta {\theta d\theta} ,\\
	\Delta A^v_w =-\sigma \eta 4\pi \bar {R}^2\int\limits_0^\theta {W_1 (\theta )d\theta } ,\\
	\Delta E^v_p =(\delta \sigma -T\frac{d\delta \sigma }{dT})(1-\eta )4\pi \bar {R}^2\int\limits_0^\theta {\theta d\theta } ,\\ 
	\Delta E^v_w =-(\sigma -T\frac{d\sigma }{dT})\eta 4\pi \bar {R}^2\int\limits_0^\theta {W_1 (\theta )d\theta } .
\end{gathered}
\end{equation}
Expressions (\ref{eq17}), like relationships (\ref{eq10}), (\ref{eq13}), and (\ref{eq14}), are valid for the case of an isothermal process. This implies that they can be used for describing experiments in the case where the characteristic time of heat transfer (removal) $\tau _Q $ is considerably shorter than the characteristic time $\tau _V $ of change in the volume of the nanoporous medium-nonwetting liquid system: $\tau _Q \ll \tau _V $. At $\tau _Q \ge \tau _V $, the temperature and, correspondingly, the quantities $\sigma $, $\delta \sigma $, $d\sigma / dT$, and $d\delta \sigma /dT$ become dependent on the time and, hence, on the degree of infiltration $\theta $. In this case, they should be introduced under the integral sign in relationships (\ref{eq14}) and (\ref{eq17}). The inequality $\tau _Q \ll \tau _V $ impose constraints on on velocity of compression system under the study of its equilibrium properties.

\section{CONDITIONS FOR THE CLOSED CYCLE}

It can be seen from relationships (\ref{eq14}) and (\ref{eq17}) that, if after the increase in pressure and the infiltration of all pores of the nanoporous medium with a liquid and the subsequent decrease in pressure and the complete defiltration, the system reverts to its original state, the following relationship should hold true: 

\begin{equation}
\label{eq18}
\begin{gathered}
(\delta \sigma -T\frac{d\delta \sigma }{dT})(1-\eta )\int\limits_0^1 {(\theta P(\theta )-\theta )d\theta } +(\sigma -T\frac{d\sigma }{dT})\eta \int\limits_0^1 {(W_1 (\theta )+W(\theta ))d\theta } =\Delta E,\\
\Delta E=0.
\end{gathered}
\end{equation}
Here, $\Delta E$ is the change in the internal energy of the system upon the transition from the initial state to the final state in the course of infiltration and defiltration. 

Expression (\ref{eq18}) relates the energy parameters of the liquid-solid and liquid-gas interfaces and the macroscopic characteristics of the porous medium, such as the porosity $\varphi $, the structure of the percolation cluster $P(\theta )$, and the quantities $W$ and $W_1 $, which determine the dependence on $\theta $ for the surface of menisci at the mouths of filled pores on the shell of an infinite cluster of filled pores in the case of infiltration and for the surface of menisci in pores on the shell of all clusters of empty pores (including the infinite cluster) in the case of defiltration. It follows from relationships (\ref{eq18}) that the absorbed energy in contraction and expansion of the system in the closed cycle is dependent on the quantities $\sigma $ and $\delta \sigma $ and equal to the total heat released upon the formation and disappearance of the liquid-solid and liquid-gas surfaces. This heat is determined by the independent quantities, namely, the derivatives of the quantities $\sigma $ and $\delta \sigma $ with respect to temperature. The integrals in expression (\ref{eq18}) account for the different paths of the system in the course of infiltration and defiltration. In the closed cycle, according to relationships (\ref{eq10}), (\ref{eq13}), (\ref{eq17}), and (\ref{eq18}), during the infiltration and defiltration, the system undergoes different sequences of equilibrium states that differ in macroscopic sets of filled pores.

In particular, the infiltration of the porous medium according to expression (\ref{eq18}) is accompanied by an increase in the number of pores that belong to the infinite cluster of filled pores and by a change in the number of menisci in pores on the shell of this cluster. The defiltration of the porous medium is accompanied by an increase in the number of pores in all clusters (including the infinite cluster of empty pores and single pores) and by a change in the number of menisci on the shell of the entire system of empty pores. As follows from relationships (\ref{eq9}), (\ref{eq10}), and (\ref{eq16}), these sequences of states depend on the porosity $\varphi $ and the number of neighbors $z$ in the system of connected pores. Therefore, in terms of the percolation theory and the model under consideration, the contact angle hysteresis is associated with different (in infiltration and defiltration) macroscopic properties of systems of filled and emptied pores that manifest themselves as different spatial distributions of the liquid in the connected pores. If the sets of macroscopic equilibrium states characterized by the distributions of filled and empty pores in the course of infiltration and defiltration were identical, the total thermal effect in the closed cycle would be equal to zero. In this case, the thermodynamic relationship (\ref{eq18}) for the closed cycle is not satisfied. 

The closed cycle and, consequently, the complete transformation of the work into the heat was observed for a number of water-hydrophobized silica gel systems \cite{7,17,18,21,22}. In particular, the complete emptying of pores after the infiltration and the subsequent defiltration with a decrease in the excess pressure to zero was observed in \cite{17,18} for the KSK-G silica gel modified by n-alkylsilane molecules ($n=8,~16$) grafted to the silica gel surface with a surface density higher than 2 nm$^{-2}$. Before the modification, the specific surface area, the pore volume, and the average pore radius for this silica gel were equal to $310\pm 20$ m$^{2}$/g, 0.95 cm$^{3}$/g, and $\bar {R}=5.2$ nm, respectively. The values of these quantities after the modification are not presented in \cite{17,18}, which complicates the analysis of the results obtained in these works. The closed cycle was also observed in \cite{21} for the water-C8W silica gel (Waters) system in which the silica gel was modified by n-alkylsilane with the chain length $n=8$, the average pore radius $\bar {R}=4.2$, and the specific pore volume of 0.53 m$^{2}$/g in the temperature range $287\div 333$ K. The authors established that the small nonoutflow ($<1${\%}) takes place only at a temperature of 278 K. For the systems containing water and the Fluka 60 C8 silica gel, as well as the Zorbax Z4, Z8, Z18, and PEP10C18 silica gels, the closed cycle was observed in \cite{7, 22}. In \cite{7}, the authors investigated the infiltration and defiltration in four porous media MCM41 with pores in the form of cylindrical channels. These media were also modified by n-alkylsilane with $n=8$ and a surface density of 2.1 nm$^{-2}$ and had the average pore radii $\bar {R}=1.3$, 1.6, 2.0, 5.4 nm. The phenomenon of nonoutflow was observed only for the porous medium with $\bar {R}=5.4$ nm, whereas the other porous media with $\bar {R}=1.3$, 1.6, 2.0 nm underwent a closed infiltration-defiltration cycle. 

For the porous medium with specified parameters of the macroscopic structure of the pore space, the closed cycle with the complete defiltration, according to relationship (\ref{eq18}), is possible only when the values of the quantities $\sigma $, $\delta \sigma $, $d\sigma /dT$ and $d\delta \sigma /dT$ fall in particular ranges. Therefore, relationship (\ref{eq18}) requires a separate detailed quantitative analysis, which will be performed elsewhere. It should be noted that this analysis, in turn, necessitates the knowledge of the values of the quantities $\sigma $, $\delta \sigma $, $d\sigma /dT$ and $d\delta \sigma /dT$; the porosity; the pore size distribution (which is changed after the modification \cite{19}); and the quantities $\Delta Q$ and $\Delta A$. Here, we restrict our consideration to the qualitative analysis of the available experimental data for which relationship (\ref{eq18}) is not satisfied and the sum of the work and the heat in the cycle is equal to the change in the internal energy of the system. In this case, the mechanical work dissipated by the system is not equal to the total heat release, as it was described in \cite{47}. The phenomenon of nonoutflow associated with the change in the internal energy is characteristic of the majority of the studied hydrophobic porous media and liquids, namely, water \cite{7, 19, 20, 23, 24, 34, 35, 36, 10,40}, aqueous solutions of salts \cite{27,28, 12}, organic compounds, ethylene glycol \cite{5}, alcohol \cite{26}, and glycerol \cite{48}, as well as systems with the liquid metal \cite{1}, Wood's alloy \cite{4}, and mercury \cite{67,68}. 

For water, the derivative involved in relationship (\ref{eq18}) has the value $d\sigma /dT=1.5 \cdot 10^{-4}$ J/m$^{2}$ K \cite{69}. As the temperature changes from 293 K to 353 K, the surface energy $\sigma$ in accordance with this value of the derivative changes by $\approx 5${\%}. Such a small change of the surface energy $\sigma $ in the system containing water and the modified Fluka 100 C8 silica gel appears to be sufficient that the closed cycle will transform into the cycle with a nonoutflow of more than 80{\%} of water with a decrease in the temperature from 353 to 293 K \cite{24}. This cycle is characterized by a change in the internal energy of the system ($\Delta E\ne 0)$. The quantity $\Delta E$ reflects both the reversible and irreversible changes in repeated infiltration-defiltration cycles. The reversible change can be associated with the adsorption of water on the modified surface \cite{19,20}. This water evaporates already at room temperature (and more rapidly at an elevated temperature), and the first cycle with nonoutflow is reproduced \cite{7,24,25}. The differential thermal analysis performed in \cite{19} demonstrated that water evaporates at a temperature $T=373$ K. The irreducible change of the internal energy $\Delta E$ is governed by the interaction of the liquid with the surface of the porous medium and depends on the maximum pressure during the infiltration \cite{25}, the temperature and time of heating of the system \cite{23}, and the procedure used for preparing the surface after the modification \cite{12}. The change of the internal energy $\Delta E$ also depends on the length ($n$) of the grafted n-alkylsilane molecule. It is worth noting that, for water and silica gels, an increase in $n$ from 4 to 18 leads to a increase $\delta \sigma $, a decrease in the nonoutflow and a decrease in the change of the internal energy $\Delta E$ \cite{7,8,16, 17, 18, 19, 20}, whereas for water and the modified medium, the closed cycle is observed for $n=1$ and the complete nonoutflow takes place for $n=8$ \cite{32}. An increase in the concentration of the NaCl or CaCl$_{2}$ salt in the aqueous solution in specific ranges leads to an increase in the quantity $\delta \sigma $ and a decrease in the change of the internal energy $\Delta E$ for the Fluka 100 C8 silica gel \cite{25,27,28,32}. A decrease in the concentration of ethylene glycol in the aqueous solution and the hydrophobic silica gel L23 results in the transition from the complete nonoutflow at a concentration $c>15${\%} to the closed cycle (the nonoutflow is less than $<$5{\%}) \cite{5}. Unlike ethylene glycol, ethanol wets the modified surface of the Fluka 100 C8 silica gel. This brings about the adsorption of ethanol and an increase in the change of the internal energy $\left| {\Delta E} \right|$ \cite{26}. 

In recent years, experimental data have been published on the infiltration of hydrophobic zeolites with water and aqueous solutions of salts \cite{35, 36, 10, 11, 39, 40, 41, 42, 43}. It has been revealed that, for systems containing water and silicalite 1 (OH), silicalite 1 (F$^-$), and ZSM-5 zeolite, the hysteresis is not observed and the pressure dependences of the volume of the system measured for the infiltration and defiltration coincide with each other. This means that such systems exhibit properties of an elastic spring without dissipation at $\Delta E=0$ in the cycle. With a cyclic change in the temperature from 358 to 318 K, these systems manifest a volume memory effect \cite{10,11}. For the water-MFI zeolite system, the dependences of the pressure on the volume and temperature during the infiltration remain unchanged when the rate of decrease in the volume changes by three orders of magnitude \cite{41}. This implies that the Laplace--Washburn and Poiseuille equations are not applicable to the system under investigation. The infiltration-defiltration hysteresis was observed for a KCl aqueous solution and the Zeolyst CBV-901 (HY) zeolite treated with SiCl$_{4}$ \cite{11}. This hysteresis depends on the nature of the anion \cite{43} and, for the Y(ZY)-zeolite, on the nature of the cation (Li, Na, K, Cs); in this case, the infiltration pressure decreases with an increase in the cation radius \cite{42}. 

In the analysis of experimental results obtained for zeolites, it is necessary to take into account that, for a channel (pore) diameter smaller than 1 nm, the liquid acquires properties of one-dimensional systems \cite{41,71, 72}, which differ qualitatively from the properties of the liquid in channels of larger sizes. This problem requires a separate analysis. 

In conclusion of the discussion of relationship (\ref{eq18}), we should note that it was derived under the assumption that $\delta R/\bar {R}<3$ in the absence of correlations in the mutual arrangement of pores of different sizes and with the use of the model of pores as a system of randomly arranged spheres. Relationship (\ref{eq18}) for the closed cycle does not contain the average pore radius $\bar {R}$; however, the value of $\bar {R}$ affects the porosity $\varphi $ and, hence, the number of neighbors $z$, the structure of the percolation cluster, and the quantities $W(\theta )$ and $W_1 (\theta )$.

\section{TEMPERATURE DEPENDENCES OF THE INFILTRATION AND DEFILTRATION PRESSURES}

Now, we analyze the signs of the derivatives $d\sigma /dT$ and $d\delta \sigma /dT$ involved in relationship (\ref{eq18}) and the integrals of the quantities $P(\theta )$ and $\left[W(\theta )+W_1 (\theta )\right]$. It follows from relation (\ref{eq10}) that the sign of the total thermal effect during infiltration is determined by signs of $\Delta Q_p $, $\Delta Q_w $ values. The sign of the thermal effect due to the formation of the liquid-solid interface $\Delta Q_p $ depends on the sign of $d\delta \sigma / dT$, which can be ascertained using known dependences of pressure at the beginning of infiltration $p_{in}$ and defiltration $p_{out} $ on temperature. The experiments carried out showed that for all investigated systems (modified silica gel -- water, aqueous solutions of salts) the pressure at the beginning of infiltration changes not much as temperature increases $dp_{in} /dT \cong 0$, while the defiltration pressure increases as temperature rises \cite{21,22,24}. Therefore, in order to determine the sign of the derivative $d\delta \sigma /dT$, we calculate the pressures required for the infiltration of one pore on the shell of the infinite cluster and the defiltration of the liquid from an arbitrary pore under the conditions $\delta A_{in} =0$ and $\delta A_{out} =0$, which follow from relationship (\ref{eq4}). By assuming that the quantities $\eta $, $\bar {R}$, and $W$ do not depend on the temperature and taking into account that the infiltration and defiltration begin in the vicinity of the corresponding percolation threshold from relationships (\ref{eq13}) and (\ref{eq15}), the derivatives $dp_{in} /dT$ and $dp_{out} /dT$ can be written in the following form: 

\begin{equation}
\label{eq19}
\frac{dp_{in} }{dT}=\frac{3}{\bar {R}}\cdot \left[ {\frac{d\delta \sigma }{dT}\cdot (1-\eta )+\frac{d\sigma }{dT}\cdot \eta W(\theta \sim \theta _c )} \right],
\end{equation}

\begin{equation}
\label{eq20}
\frac{dp_{out} }{dT}=\frac{3}{\bar {R}}\cdot \left[ {\frac{d\delta \sigma }{dT}\cdot (1-\eta )-\frac{d\sigma }{dT}\eta W_1 (\theta \sim 1-\theta _c )} \right].
\end{equation}
It follows from eq. (\ref{eq12}),(\ref{eq16}) that $W(\theta \sim \theta _c )>0$, $W_1 (\theta \sim 1-\theta _c )>0$, $\theta _c =0.18$. Since $dp_{in} /dT\cong 0$, it follows from eq. (\ref{eq19}), (\ref{eq20}) that

\begin{equation}
\label{eq21}
\frac{d\delta \sigma }{dT}\cdot (1-\eta )\sim -\frac{d\sigma }{dT}\cdot \eta W(\theta \sim \theta { }_c).
\end{equation}
Hence, the sign of $d\delta \sigma /dT$ is opposite to the sign of $d\sigma /dT$. The coefficient of the surface tension of the liquid-gas interface decreases as temperature increases, and vanishes at critical point so that $d\sigma /dT<0$ \cite{70}. For water, value $d\sigma /dT$ is $-1.5 \cdot 10^{-4}$ J/m$^{2}$K \cite{70}. In this case, from eq. (\ref{eq20}) find that for pressure at the beginning of infiltration $dp_{out} /dT>0$, which corresponds to experimental data \cite{21,24}.

Since the probability that a pore belongs to an infinite cluster is $P(\theta )\le 1$ and $\int\limits_0^1 {(\theta P(\theta )-\theta )d\theta } \le 0$, it is necessary to analyze the dependences $W(\theta )$ and $W_1 (\theta )$ in order to determine the sign of the thermal effect in the course of the infiltration and defiltration. The behavior of the quantities $W_1 (\theta )$ and $W(\theta )$ and, hence, the sign of the integral $\int\limits_0^1 {(W_1 (\theta )+W(\theta ))d\theta } $ depend substantially on the porosity $\varphi $. The figures shows the dependences $W(\theta )$ and $W_1 (\theta )$ calculated from relationships (\ref{eq12}) and (\ref{eq16}) for different values of the porosity $\varphi $.

\begin{figure}[h]
\center{\includegraphics[width=0.5\linewidth]{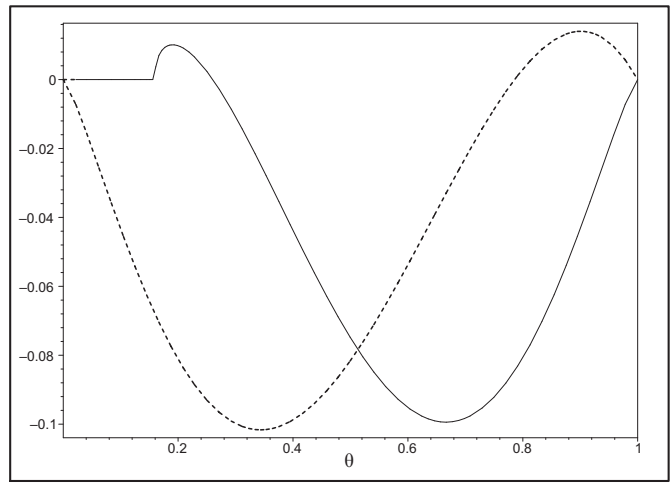}}
\caption{The dependences $W(\theta )$ (solid line) and $W_1 (1-\theta )$ (dashed line) for porosity $\varphi =0.25$}
\label{fig2}
\end{figure}

\begin{figure}[h]
\center{\includegraphics[width=0.5\linewidth]{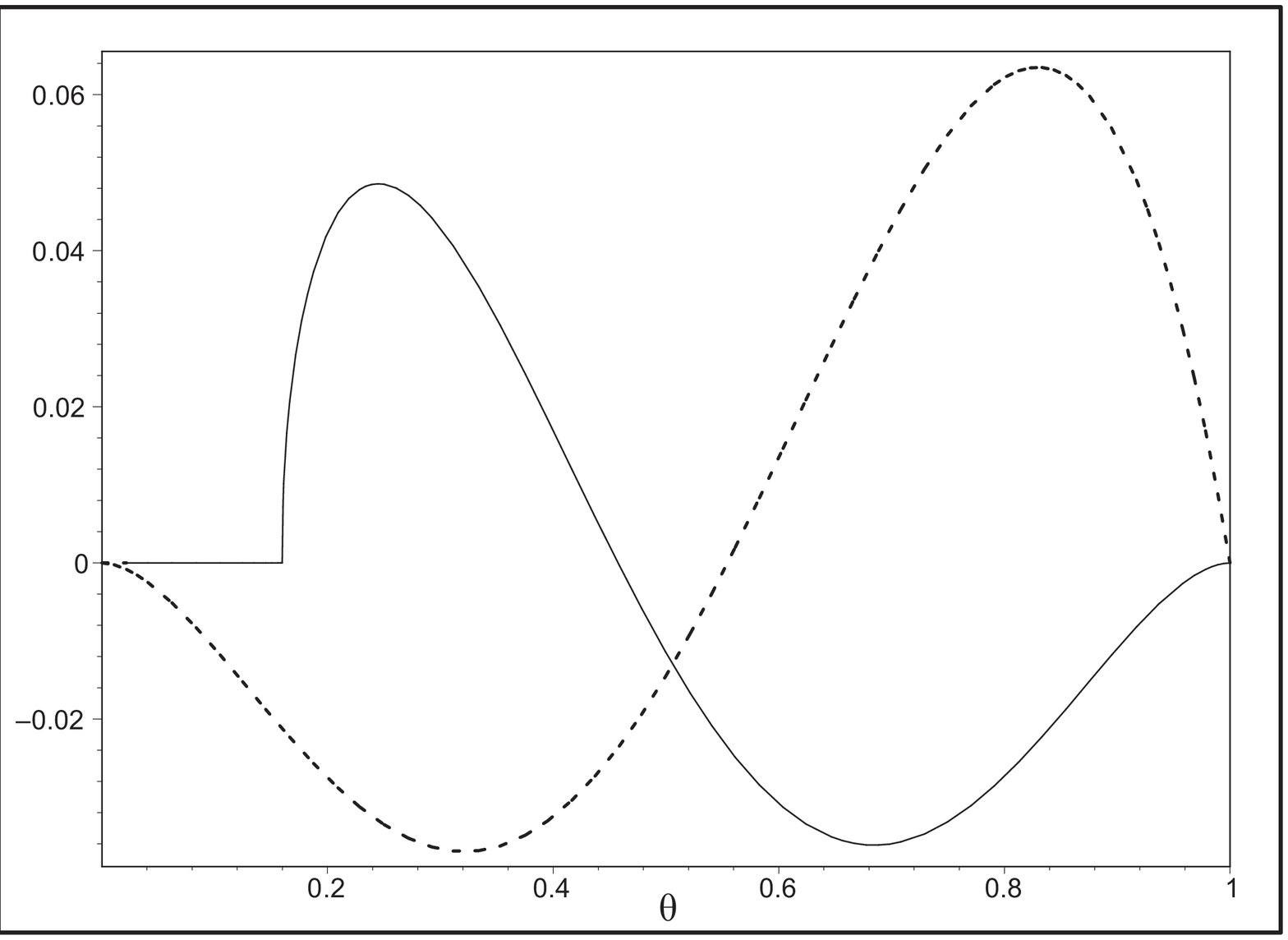}}
\caption{The dependences $W(\theta )$ (solid line) and $W_1 (1-\theta )$ (dashed line) for porosity $\varphi =0.3$}
\label{fig3}
\end{figure}

\begin{figure}[h]
\center{\includegraphics[width=0.5\linewidth]{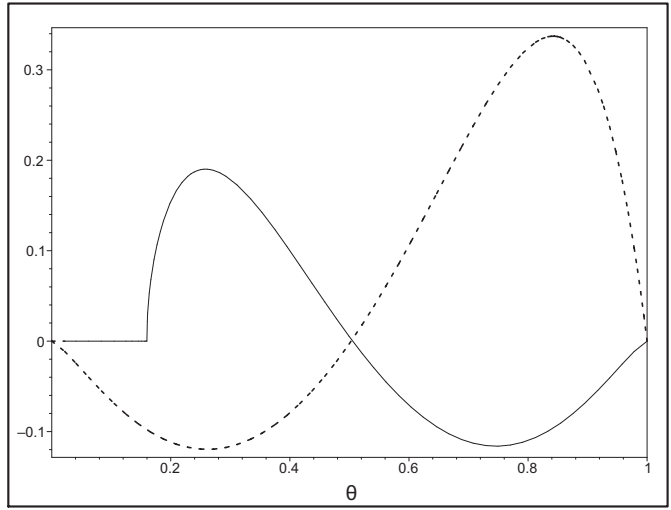}}
\caption{The dependences $W(\theta )$ (solid line) and $W_1 (1-\theta )$ (dashed line) for porosity $\varphi =0.6$}
\label{fig4}
\end{figure}

It follows from the figures that the $W(\theta )$ and $W_1 (\theta )$ functions nontrivially depend on porosity. Since the filling of the porous medium begins only with the formation of the infinite cluster, $W(\theta )=0$ for $\theta <\theta _c =0.18$. At low porosity $\varphi \sim 0.22$ the structure of pores being filled is close to the fractal structure of the low-density infinite cluster near the percolation threshold, so that the growth of its surface is compensated by the decrease in the difference between the number of menisci in final and initial states of the filled pore at $\theta \le 0.3$, since in concordance with (\ref{eq9}) there are few nearest neighbors of this pore $\bar {z}<3$ (Fig.~\ref{fig2}). During further infiltration only the reduction in number of emerging menisci occurs (Fig.~\ref{fig2}), which reaches minimum at $\theta \approx 0.68$. If $\theta =1$ the value $W(1)$ is zero, since the filled porous medium meniscus is absent.

For media with high porosity $\varphi >0.3$ in which the number of nearest neighbours is $\bar {z}>3$, as the degree of filling $\theta \ge \theta _c $ increases, the infinite cluster of filled pores grows, accompanied by the growth of its surface and increase in the number of contacts of an empty pore with its neighbours on the shell of the infinite cluster filled with liquid. It leads to decrease in the difference between the number of menisci in final and initial states of the filled pore, so that as $\theta $ increases the $W(\theta )$ value reaches maximum and then decreases (Fig.~\ref{fig3},~\ref{fig4}). In accordance with (\ref{eq12}) it will continue until decrease in the difference between the number of menisci in final and initial states of the filled pore compensate the growth of area of the infinite cluster, which will lead to vanishing of $W(\theta )$ at $\theta \sim 0.5$ (Fig.~\ref{fig2},~\ref{fig3}). The further growth of the degree of filling leads to further decrease in the difference between the number of menisci in final and initial states of the filled pore while the growth of the surface of the infinite cluster of filled pores slows down. As a result $W(\theta )$ reaches maximum at $\theta \sim 0.7$. The surface of the infinite cluster of filled pores will vanish at $\theta =1$, which will lead to vanishing of $W(\ref{eq1})$ (Fig.~\ref{fig3},~\ref{fig4}). The dependence of the $W_1 (1-\theta )$ value, which determines defiltration, is similar to the dependence of $W(\theta )$, which determines infiltration. The difference in behavior of $W(\theta )$ and $W_1 (1-\theta )$ is connected with the fact that liquid defiltration from the porous medium does not require the percolation cluster formation. Figures \ref{fig2}-\ref{fig4} show that $W(\theta \sim \theta _c )>0$, $W_1 (\theta \sim 1-\theta _c )>0$. 

The calculation of the integral $\int\limits_0^1 {(W_1 (\theta )+W(\theta ))d\theta } $ shows that, with an increase in the porosity, this integral increases from negative values at $\varphi =0.25$ to positive values at $\varphi =0.6$ and vanishes at $\varphi =0.3$. 

\begin{figure}[h]
\center{\includegraphics[width=0.5\linewidth]{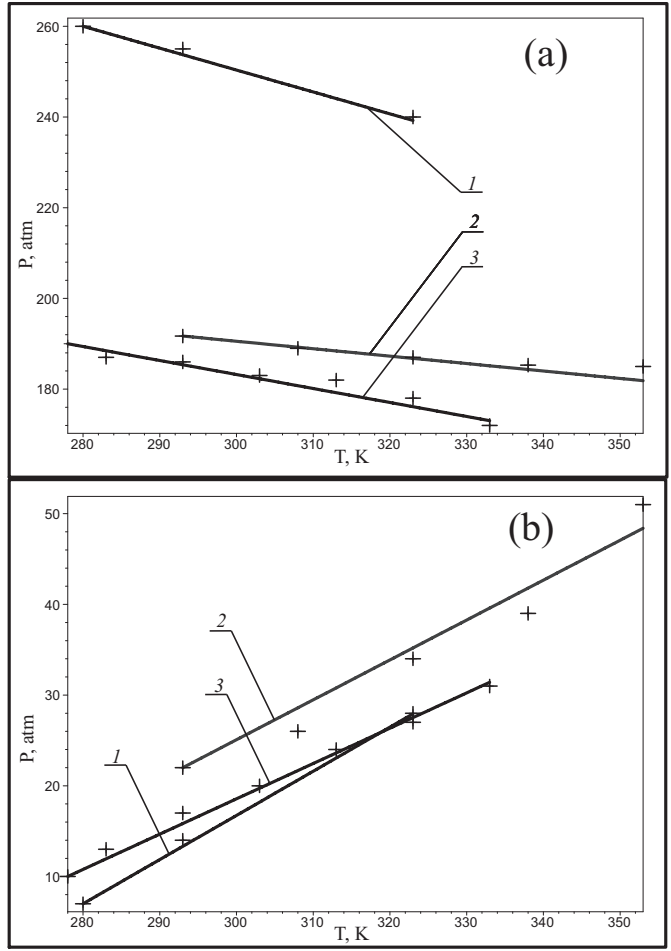}}
\caption{Dependences (a) $p_{in} (T)$and (b) $p_{out} (T)$ calculated for Libersorb 23 (curves 1), Fluka 100 C8 (curves 2), and C8W (curves 3) infiltrated with water according to the calculations from relationships (\ref{eq13}) and (\ref{eq15}). Points are the experimental data taken from \cite{24,46}}
\label{fig5}
\end{figure}

Expressions (\ref{eq13}), (\ref{eq15}), (\ref{eq19}), (\ref{eq20}) and conditions $\delta A_{in} =0$ and $\delta A_{out} =0$ allow one to calculate the temperature dependences of the infiltration pressure $p_{in} (T)$ and the defiltration pressure $p_{out} (T)$. Figure~\ref{fig5} shows the experimental data for silica libersorb 23 (silica gel KSK-G with the modification of 8-tier alkynesilan-C8), Fluca100 C8 \cite{24}, and S8W \cite{21} filled by water and calculated according to the considered correlation effects. Experimental data within the measurement error are described by linear dependences with different slopes. With increasing temperature, the pressure of infiltration decreases at about 10{\%} and the defiltration pressure increases in the times for all the porous media. For these environments values of pressure $p_{in} $ and $p_{out} $ also differ at initial temperature. In concordance with relationships (\ref{eq13}) and (\ref{eq15}), dependences $p_{in} (T)$ and $p_{out} (T)$ are described by the quantities $W(\theta =\theta _0 >\theta _c )$, $W_1 (\theta =1-\theta _0 )$, and by dependences $\sigma (T)$ and $\delta \sigma (T)$. The dependence of $\sigma (T)$ is known \cite{70}. The value $\delta \sigma $ in our experiments for libersorb 23 was $\delta \sigma =22\pm 1$ J/m$^{2}$. The same value was taken for the other two porous media because of their similar modifications. The value $d\delta \sigma /dT$ was calculated from the condition (\ref{eq21}). Quantities $W$ and $W_1$ calculated according to expressions (\ref{eq9}), (\ref{eq12}), (\ref{eq16}). The porosities estimated from the specific pore volumes according to the data taken from \cite{21,24} were as follows: $\varphi =0.33$ for Libersorb 23, $\varphi =0.46$ for Fluka 100, and $\varphi =0.53$ for C8W. The quantities $W(\theta =\theta _0 >\theta _c )$ and $W_1 (\theta =1-\theta _1,\theta _1 <<1)$ corresponded to the maximum values in the curves $W(\theta )$ and $W_1 (1-\theta )$ for $\varphi =0.33$, $\varphi =0.46$, and $\varphi =0.53$, respectively. 

As is also seen from Fig.~\ref{fig5}, the calculated temperature dependences of the infiltration and defiltration pressures for the porous medium Libersorb 23 C8, Fluka 100 C8, and C8W infiltrated with water satisfactorily fit the experimental data. 

It follows from relationships (\ref{eq13}) and (\ref{eq15}) that, in the framework of the proposed model, the pressures of the beginning of the infiltration and defiltration $p_{in} $ and $p_{out} $ are proportional to $1/\bar {R}$ if the quantities $\eta $, $W$, and $W_1 $ are independent of $\bar {R}$. This is consistent with the known experimental data \cite{6,7,19,46,55,57}. Such dependence is a consequence of the lack of correlations in the mutual arrangement of pores of different sizes in a model of randomly arranged spheres with a narrow pore size distribution $\delta R/R<3$, characteristic for silica gels. For the porous medium MSM 41 more than the strong dependence of the average radius \cite{6,7} due to the peculiarities of the fluctuation formation of the nucleus in a cylindrical channel (see (\ref{eq3})).

\section{THERMAL EFFECT}

Expressions (\ref{eq10}) and (\ref{eq12}) allow one to calculate the thermal effects observed during the infiltration of a porous medium with a nonwetting liquid in different cases as functions of the porosity and surface energies of the liquid and the porous medium. 

It was found in \cite{46} that when one-third volume of the porous medium from Sigma-Aldrich was filled by water, the temperature did not increase within the limits of error ($\le 0.1$K). A maximum value of the elastic energy in performed tests was about 1 J. Authors \cite{46} estimated the possible increase of temperature, in condition that the work ($A$) of filling is to raise the temperature. In experiments \cite{46} at defined value $A$ as 2.9 J, the temperature increase should be $\Delta T = 0.8$K \cite{46}. Figure~\ref{fig5} shows the calculated (from relationships (\ref{eq10}) and (\ref{eq17})) dependences of the thermal effect on the degree of infiltration of the porous medium (in relative units) due to the formation of the liquid-porous medium surface $\Delta Q_p (\theta )/A$ (Fig.~\ref{fig5}a, lower curve) and menisci $\Delta Q_w (\theta )/A$ (Fig.~\ref{fig5}a, upper curve) and the corresponding dependence of the total effect $\Delta Q/A=(\Delta Q_p +\Delta Q_w)/A$ (Fig.~\ref{fig5}b). In accordance with (\ref{eq10}), (\ref{eq12}), the value of the thermal effect depends on the value of the integrals appearing in these relations, which, (see (\ref{eq9}),(\ref{eq12}) and (\ref{eq16})) depend on the porosity. Estimates show that in these experiments the porosity of silica gel after modification was $\varphi \approx 0.68$. The parameter $T \cdot d\delta \sigma /dT$ was calculated from the condition (\ref{eq21}).

\begin{figure}[h]
\center{\includegraphics[width=0.5\linewidth]{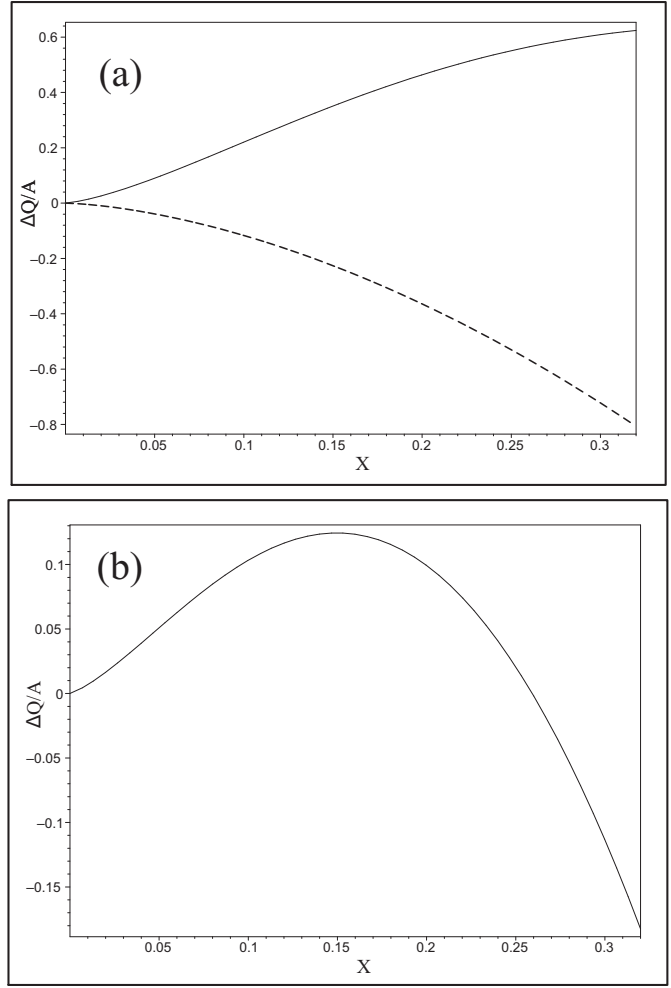}}
\caption{Dependences of the thermal effect (in relative units) on the degree of infiltration $\theta $ of the porous medium ($\varphi \approx 0.68)$ with water: (a) the contributions due to the formation of the liquid-porous medium surface $\Delta Q_p (\theta )/A$ (lower curve) and menisci $\Delta Q_w (\theta )/A$ (upper curve) and (b) the total thermal effect $\Delta Q/A=(\Delta Q_p +\Delta Q_w )/A$, $A$=2.9 J}
\label{fig6}
\end{figure}

Note that $W(\theta )$ is equal to zero at $\theta <\theta_c$. Therefore, in Fig.~\ref{fig6} for comparison with experimental data, degree of infiltration is delayed on the horizontal axis, $X$ shifted by an amount $\theta _c =0.18$ so that the value $\theta=0$ corresponds $\theta =\theta _c $ in the equation (\ref{eq12}).

It can be seen from Fig.~\ref{fig6} that the thermal effect associated with the infiltration of the porous medium with the above parameters is small: at the maximum, it reaches $\sim 0.15A$ for the degree of infiltration $X\sim 0.17$. For $X\approx 0.27$, the thermal effect vanishes. This is explained by the different origins of the contributions from the menisci and the pore surface to the total thermal effect, so that each contribution is one order of magnitude larger than the total thermal effect. 

Thus, in the performed experiments \cite{47}, upon the infiltration of the porous medium with water (the heat capacity is 4.2 J/g K \cite{70}) for the heat release $Q\sim 0.15A=0.45$ J, the maximum increase in the temperature is $\Delta T\sim 0.2$K. In this case, it should be expected that, according to relationship (\ref{eq18}), the change of the internal energy of the system $\Delta E$ in the performed experiments upon the transition from the initial state to the final state during the infiltration and defiltration differs from zero and is comparable in the order of magnitude to the work expended for infiltrating the porous medium: $\Delta E\sim A$. This internal energy is equal to the difference between the energy of menisci in the pores on the shell of the infinite cluster of filled pores and the surface energy of the porous medium-liquid interface in the filled pores. The positive energy of system compression  also give contribution to $\Delta E$.

\begin{figure}[h]
\center{\includegraphics[width=0.5\linewidth]{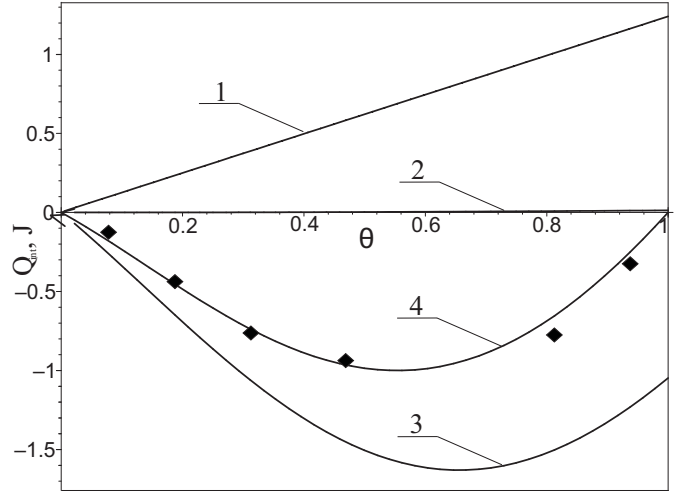}}
\caption{Dependences of the heat release $Q = \Delta Q(\theta )/Q_0$ on the degree of infiltration according to the measurements performed in \cite{2} (points) and calculations from equations (\ref{eq9}), (\ref{eq12}) with allowance for the compressibility of the porous medium with $\varphi \sim 0.22$}
\label{fig7}
\end{figure}

Relationships (\ref{eq9}) and (\ref{eq12}) make it possible to compare the values of the heat release observed in the experiments on the infiltration of the KSK-G (modification C16) porous medium with water \cite{2}. According to the estimates, the medium porosity in the experiments performed in \cite{2} is $\varphi \approx 0.22$. For this medium, we calculated the thermal effect associated with the infiltration. Figure~\ref{fig7} shows experimental data and the dependence of the thermal effect (in relative units) $\Delta Q(\theta )/ \left| Q_0 \right|$ (where $ \left|Q_0 \right|$ is the maximum heat release upon the infiltration of the porous medium with water; according to \cite{2}, $Q_0 = 4$ J/g) calculated from relationships (\ref{eq9}) and (\ref{eq12}). The elastic energy of compression of the porous medium and water is determined by water $\Delta Q_u = -T \Delta V \cdot dK/dT $, where $\Delta V$ is the specific decrease in the volume of the water-porous medium system and $K= 0.166$ mN/mK is the compressibility of water \cite{2}. The maximum value of $\left| \Delta Q_u \right|$ is approximately equal to 6 J/g.

Figure~\ref{fig7} also presents the components of the thermal effect measured in \cite{2}: the dependence of the heat release associated with the compressibility of the system (curve 1) and the dependence of the heat release $\Delta Q_p $ due to the change in the surface energy of pores (curve 2) and menisci $\Delta Q_w $ (curve 3). It should be noted that the porous medium studied in \cite{2} has a porosity $\varphi \sim 0.22$ and, hence, its system of pores is located in the vicinity of the percolation threshold $\varphi _c =0.18$. Consequently, the infinite cluster of empty pores, which is required for the infiltration of the porous medium, is very sparse \cite{10} and the probability (involved in relationship (\ref{eq10})) that a pore belongs to an infinite cluster is low $P \ll 1$, therefore term contribution $\Delta Q_p $ to the thermal effect is small. In this case, it follows from relationship (\ref{eq10}) that, in the infiltration of this medium, the decisive role will be played by the second term in expression (\ref{eq10}), which containing integral from $W(\theta )$ and corresponds to the contribution of menisci to the thermal effects ($\Delta Q_w <0$). It is this circumstance that is responsible for the unusual thermodynamic properties of the system used in \cite{2}. It can be seen from figure~\ref{fig7} that calculated dependences of total thermal effect is in good agreement with the experimental data. 

\begin{figure}[h]
\center{\includegraphics[width=0.5\linewidth]{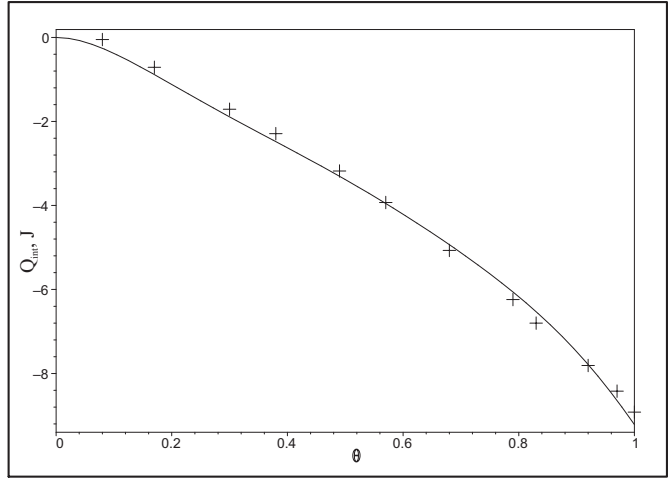}}
\caption{Dependence of the heat release on the degree of infiltration according to the measurement performed in \cite{50} (points) for porous medium KSK-G and calculations from equations (\ref{eq9})-(\ref{eq12}) with allowance for the compressibility of the porous medium $\varphi \sim 0.4$}
\label{fig8}
\end{figure}

\begin{figure}[h]
\center{\includegraphics[width=0.5\linewidth]{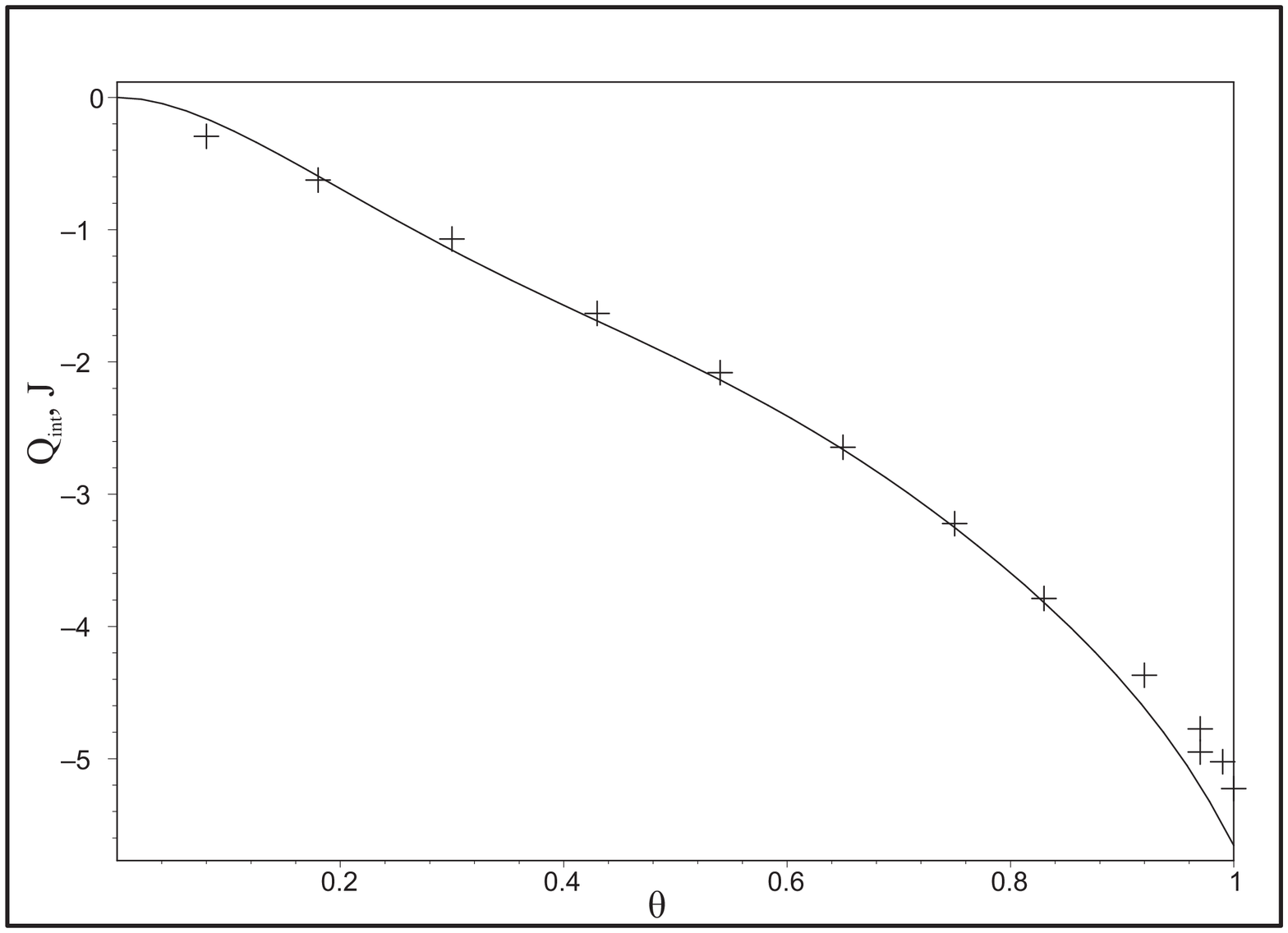}}
\caption{Dependence of the heat release on the degree of infiltration according to the measurement performed in \cite{49} (points) for porous medium PEP300 C18 and calculations from equations (\ref{eq9})-(\ref{eq12}) with allowance for the compressibility of the porous medium $\varphi \sim 0.58$}
\label{fig9}
\end{figure}

\begin{figure}[h]
\center{\includegraphics[width=0.5\linewidth]{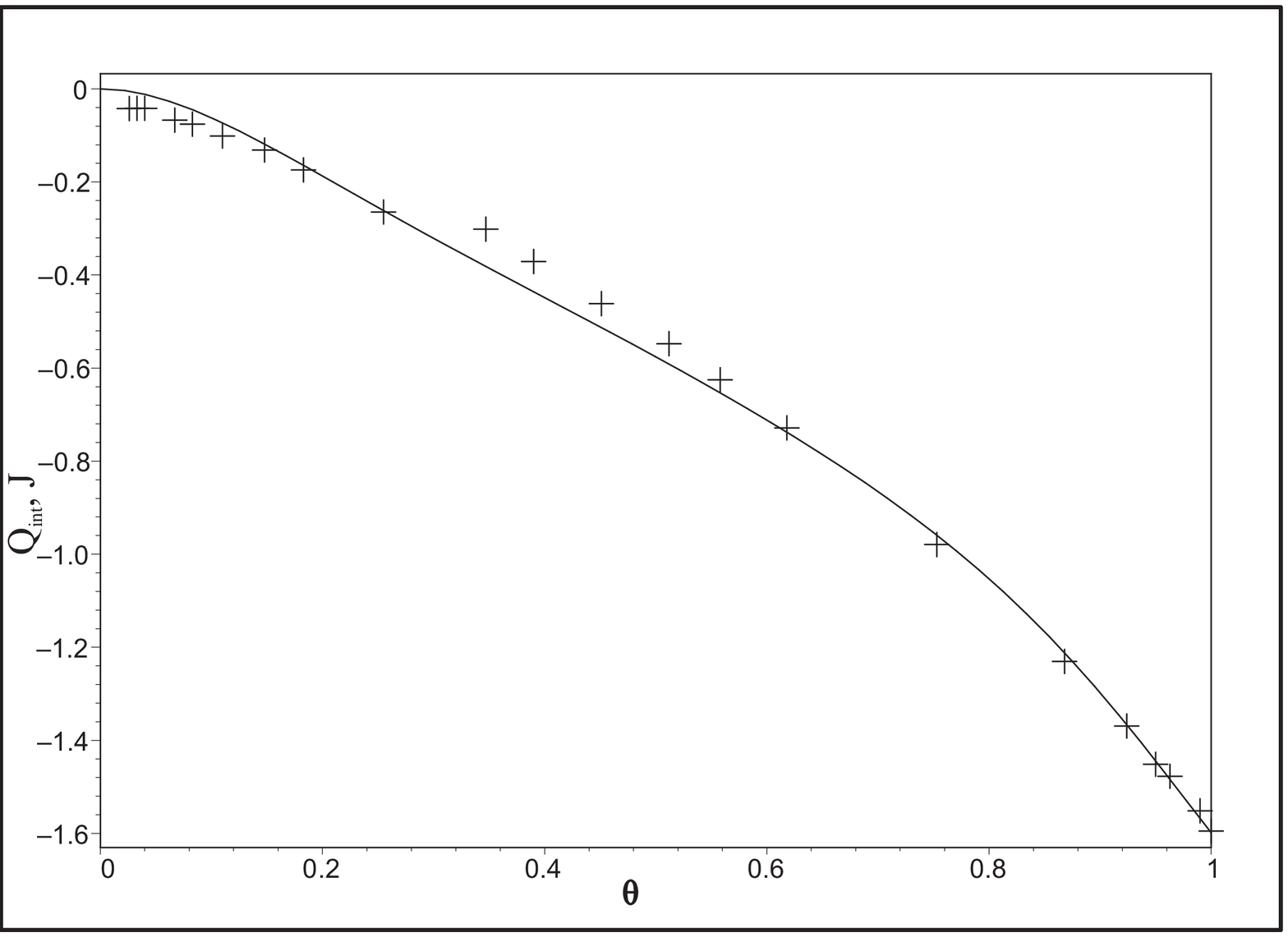}}
\caption{Dependence of the heat release on the degree of infiltration according to the measurement performed in \cite{49} (points) for porous medium PEP100 C18 and calculations from equations (\ref{eq9})-(\ref{eq12}) with allowance for the compressibility of the porous medium $\varphi \sim 0.65$}
\label{fig10}
\end{figure}

Relationships (\ref{eq9})-(\ref{eq12}) make it possible to explain the thermal effect observed in the experiments \cite{49,59} on KSK-G (modification C16) with porosity  $\varphi \sim 0.4$, on PEP100 (modification C18) and PEP300 (modification C18) with porosity $\varphi \sim $0.65 and $\varphi \sim $0.58 correspondingly. The average radii of the investigated porous medium are 6.5 nm, 5 nm and 15 nm for the KSK-G (C16), PEP100 and PEP300, respectively. The infiltration of the KSK-G silica gel was carried out at 308$^{\circ}$ K and the infiltration of the PEP100 and PEP300 were carried out at 298$^{\circ}$ K. Figures~\ref{fig8}-\ref{fig10} show experimental data \cite{49,59} and dependences of the thermal effect (in relative units) calculated from relationships (\ref{eq9})-(\ref{eq12}) for the KSK-G (C16), PEP100(C18) and PEP 300(C18). The filled volume was normalized and transformed to the degree of filling in calculations for dependences from relations (\ref{eq9})-(\ref{eq12}). The elastic energy was estimated from the experimental data of dependence of pressure on the volume, taking into account the compressibility of the container and liquid. The parameter $T \cdot d\delta \sigma /dT$ was calculated from the equation (\ref{eq21}) as previously.

Thus, for the investigated porous media with $\varphi =0.4$, $\varphi \sim 0.58$ and $\varphi \sim 0.65$ value of the thermal effect is negative and decreases with increasing the degree of filling, reaching a minimum at complete infiltration. It is indicates, that menisci give a significant contribution to thermal effects observed in \cite{49,50}. It can be seen from figures~\ref{fig7}-\ref{fig10} that calculated dependences of total thermal effect is in good agreement with the experimental data. 

The measurements performed in \cite{19,9} during multi-cycle infiltration-defiltration process (number of cycles $\sim 1200$) demonstrated that the increase in temperature in the systems containing hydrophobized silica gels and water per cycle is less than $<10^{-2 }$K. The analysis of the experimental data reported in \cite{19} showed that, in the course of the cyclic infiltration-defiltration process, the hysteresis and, hence, the heat release in the first cycle are two times larger than those observed in the 1200th cycle. Thus, in the experiments conducted in \cite{19}, there occurs a partial nonoutflow of the liquid from the porous medium, which leads to changes in parameters and characteristics of the porous medium, such as the porosity, the average number of nearest neighbors, and the interfacial energy. In this case, it is necessary to perform a detailed analysis of the changes in the parameters of the porous medium in each cycle with inclusion of the thermal diffusivity of the porous medium and water in terms of the above-derived relationships, which is beyond the scope of our present work. 

\section{CONCLUSIONS}

Thus, in this work, we have established the relation between the energy properties of the interface and the macroscopic properties of the pore space in a disordered porous medium and proposed the mechanism of energy absorption during the infiltration of the nanoporous medium with a nonwetting liquid. It turned out that thermal effects observed during the infiltration of the porous medium and the defiltration of the liquid from it can be positive and negative depending on the porosity and, hence, on the number of nearest neighbors of filled pores. 

The proposed mechanism is based on the inclusion of correlation effects during percolation infiltration of an infinite disordered porous medium. It is assumed that the infiltration of the porous medium is the result of growth of a percolation cluster consisting of filled pores through the attachment of empty pores (accessible to infiltration) to the shell of this cluster. Menisci appear and disappear in pores on the shell of the percolation cluster in the course of its growth, and these processes depend on the degree of infiltration.

The above analysis is based on the representation of a system of pores in a porous medium in terms of the model of randomly arranged spheres. However, in this model, the correlations in the mutual arrangement of pores of different sizes are ignored and, hence, it is impossible to adequately describe the effects of blocking of the liquid in pores with large radii that are surrounded by pores with smaller radii. Therefore, in the framework of the model of randomly arranged spheres, we can describe the infiltration of disordered porous media only with a narrow distribution of pores over the sizes $(\delta R) \ll \bar {R}$. We assumed that pores in the porous medium have a regular (either spherical or cylindrical) shape, ignored the coordinate dependence of the surface energy of a pore, and operated actually with the average values of this energy on the surface of the corresponding pore. 

One of the consequences of the proposed mechanism is a condition that determines the class of systems for which a closed infiltration-defiltration cycle can exist. According to this condition, the initial and final states of the system coincide, the change of the internal energy is equal to zero, and the work expended for infiltrating the porous medium, which is determined by the area of the hysteresis loop, is equal to the thermal effect. Another consequence of the proposed approach is the dependence of the effective contact angle on the degree of infiltration of the porous medium with a liquid (see relationships (\ref{eq13}) and (\ref{eq15})). This dependence results from different paths on the way to the final state of the disordered porous medium during the infiltration (defiltration) with a liquid. 

Thus, the proposed approach makes it possible within a unified context to describe the temperature dependences of the infiltration and defiltration pressures for a porous medium with a disordered structure and the thermal effects associated with the absorption of the energy by ``disordered porous medium-nonwetting liquid'' systems.

\textbf{ACKNOWLEDGMENTS}

In conclusion, it is our duty to thank Professor Yu Qiao for additional information on the experiments described in the publications of his research group.

This work was supported by the analytical departmental target program ``Development of scientific potential of higher school for 2009-2010'' and the Federal target program ``Scientific and scientific-educational personnel of innovative Russia''

\end{document}